%% file: main.tex
\documentclass[submission,copyright,creativecommons]{eptcs}
\providecommand{\event}{PLACES 2019}
\usepackage{breakurl}
\usepackage{graphicx}
\usepackage{scribble}
\usepackage{latexsym}
\usepackage{listings}
\usepackage{tikz}
\usepackage{color,soul}
\usetikzlibrary{arrows,automata,calc,fit,positioning,shapes}
\usepackage{xcolor}
\usepackage[draft]{fixme}
\fxsetup{theme=color}
\FXRegisterAuthor{NN}{aNN}{NN}
\FXRegisterAuthor{JK}{aJK}{JK}
\FXRegisterAuthor{NY}{aNY}{NY}
\usepackage{underscore}
\usepackage{subcaption}
\usepackage{wrapfig}

\input{macros-haskell}

\title{Multiparty Session Type-safe Web Development\\with Static Linearity}
\def\titlerunning{MPST-safe Web Development with Static Linearity}
\author{%
  Jonathan King \institute{Imperial College London \& Habito}
\and%
  Nicholas Ng \institute{Imperial College London}
\and%
  Nobuko Yoshida \institute{Imperial College London}
}
\def\authorrunning{King, Ng, Yoshida}

\begin{document}
\maketitle
\input{abstract}

\input{introduction}

\input{scribble}

\input{encoding}
\input{impl}
\input{casestudy}

\input{related}

\section*{Acknowledgements}
We thank the anonymous reviewers for their feedback.
This work is partially supported by
EPSRC projects
EP/K034413/1,
EP/K011715/1,
EP/L00058X/1,
EP/N027833/1,
and
EP/N028201/1.

\bibliographystyle{eptcs}
\bibliography{main}

%\appendix
%\input{appendix}

\end{document}

%% file: macros-haskell.tex
\usepackage{listings}
\lstdefinestyle{hask}{
  morekeywords={foreign},
  morekeywords=[2]{Cons,Nil},
  morekeywords=[3]{Initial,Terminal,Connect,Disconnect,Send,Recv,Receive,Branch,Select},
  morekeywords=[4]{send,connect,session,choice,receive,branch,connect,disconnect},% combinators
  basicstyle=\ttfamily,
  keywordstyle={\bfseries\color{blue}},
  keywordstyle=[2]{\color{green!50!black}},
  keywordstyle=[3]{\itshape},
  keywordstyle=[4]{\color{violet!80!black}},
  stringstyle={\color{purple}},
  numberstyle=\sffamily\small,
  mathescape=true,
  showstringspaces=false,
  captionpos=b,
  escapeinside={--/*}{*/--},
}
\newcommand{\lsthask}{\lstinline[language=Haskell,style=hask]}

\newcommand{\myparagraph}[1]{\noindent{\bf #1}\ }

%% file: abstract.tex
\begin{abstract}
  Modern web applications can now offer desktop-like experiences from within the browser, thanks to technologies such as
  WebSockets, which enable low-latency duplex communication
  between the browser and the server.
  While these advances are great for the user experience,
  they represent a new responsibility for web developers
  who now need to manage and verify the correctness of more complex  and potentially stateful
  interactions in their application.

  In this paper, we present a technique for developing interactive web applications that are statically guaranteed to communicate following a given protocol.
First, the global interaction protocol is described in the Scribble protocol language -- based on multiparty session types. Scribble
  protocols are checked for well-formedness, and then each role is projected to a Finite State Machine representing the structure of communication from the perspective of the role.  We use source code generation and a novel type-level encoding of FSMs using multi-parameter type classes to leverage the type system of the target language and guarantee only programs that communicate following the protocol will type check.
  %only programs that communicate following the protocol will type check.
  %
  
  Our work targets PureScript -- a functional language that compiles to JavaScript -- which crucially has an expressive enough type system to provide \emph{static} linearity guarantees.  
  We demonstrate the effectiveness of our approach through a
  web-based Battleship game where communication is performed through WebSocket connections.
  
\end{abstract}

%% file: introduction.tex
\section{Introduction}
A common trait amongst modern JavaScript-based
interactive web apps is continuous stateful communication between the
clients and the servers to keep the interactions responsive. This is in stark
contrast to more traditional web pages, the related REST
architecture~\cite{thesis:fielding}, where a stateless HTTP request-response is
sufficient to retrieve static web pages and their associated resources from
the servers.
These usecases led to the emergence of advanced communication transports over
HTTP connections such as WebSockets~\cite{rfc6455} protocol or the
WebRTC~\cite{web:webrtc} project,
which provide web apps with full-duplex communication channels (between
browser-server and browser-browser respectively) from within the web browser. In
addition to the performance improvements by reducing
connections per HTTP request to a single persistent connection,
they enable structured, bidirectional communication patterns not possible
or convenient with only stateless HTTP connections.

As the complexity of interactions in a web app approaches that of a networked
desktop application, it becomes increasingly important to ensure that the web app
is free of communication errors that may lead to its incorrect execution.
Hence the implementation should be verified against a protocol
specification.

Consider a simple turn-based board game \textit{Battleship} between two
players. Each player starts the game by placing battle ships (contiguous
rectangles) on a 2D grid, where the ships configuration is not revealed to
the opponent. Players then take turns to guess the coordinate of opponent's
ships, where the opponent respond if it is a \textit{hit} or a \textit{miss},
and the game continues until all ships of one player have been sunk.
We will use this game as our running example in the rest of the paper.

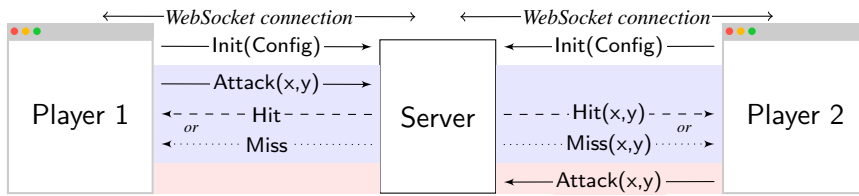
\begin{wrapfigure}{r}{0.75\textwidth}
  \centering%
  \input{fig-battleship}
  \caption{An architecture and message pattern of web-based \textit{Battleship}.}
  \label{fig:battleship-arch}
\end{wrapfigure}

A web-based implementation of the game may use the architecture depicted in
\figurename~\ref{fig:battleship-arch}. The players' clients and the game server
are connected by bidirectional WebSocket connections. Each connection
follows a predefined structured communication protocol, where the sequence
and the label of the messages passed between the participants are deterministic.
\figurename~\ref{fig:battleship-arch} also 
shows a snapshot of the messaging pattern between the
participants, which is divided into three phases: initialisation, Player 1's
turn, and Player 2's turn. During initialisation, both players send an
\textsf{Init(Config)} message to the server with their ship configurations.
The game then enters the second phase where Player 1 indicates the coordinate
to attack with a \textsf{Attack(x,y)} message, followed by a reply from the
server which is either a \textsf{Hit} or a \textsf{Miss} message. At the same
time, the Server forwards the \textsf{(x,y)} to Player 2 as a \textsf{Hit(x,y)} or
a \textsf{Miss(x,y)} message respectively. Finally, the roles of
Player 1 and 2 are reversed, and the game alternates between Player 1 and
2, until a winner can be decided.

While implementations of the game may use different user interfaces (e.g.~web
forms, graphical with HTML5 canvas), a correct implementation of the game
client should conform to the aforementioned predefined communication protocol
for the communication aspects of the game.
We use 
\emph{Multiparty Session Types} (MPST)~\cite{HYC2016} 
to specify and verify communication protocols. 
%The approach has found broad
%applications in static verification of mainstream typed languages such as Java
%and Scala for implementing communication-safe distributed programs.
%%
%However, o
Since our target endpoint language, JavaScript, is dynamically typed, 
to apply the code generation methodology from MPST,  
we could use a statically typed language
and cross compile the language to JavaScript.
Tools such as OCaml's \texttt{Js\_of\_ocaml}~\cite{vouillon-balat:spe2013,ocsigen:www2012}
compiler or Haskell's GHCJS~\cite{gh:ghcjs}
compiles the respective language into JavaScript, but the generated binaries are large in
size and these languages have their own runtimes on top of JavaScript which can complicate integration with existing JavaScript code.
Alternatively, we can apply MPST to typed languages that are designed for
JavaScript generation, examples include
Microsoft's TypeScript~\cite{web:typescript}
% -- a gradually typed superset of the ES6 JavaScript specification,
or Google's Dart~\cite{web:dart}. 
% -- a high performance custom runtime for JavaScript.
Their type systems are, however, fairly basic compared to 
functional programming languages, which restricts the static guarantees we can
provide. 

In this work we use \emph{PureScript} \cite{web:purescript}, 
a pure functional
language inspired by Haskell, which compiles to human readable JavaScript and doesn't have a runtime. It has good library support for web development and a library for cooperatively scheduling asynchronous effects, of which further details are given in \S~\ref{sec:implementation}.\\[1mm]
\myparagraph{Contributions}
This paper presents a type-safe web application development work flow following
the MPST framework: {\bf (1)}
A first {\em multiparty} session-based code generation work flow
targeting interactive web applications; 
{\bf (2)}
 A novel encoding of Endpoint FSMs~\cite{HY2016} 
using multi-parameter type classes; and 
{\bf (3)}
A lightweight session runtime using the encoding which also statically
    prevents non-linear usage of communication channels.  

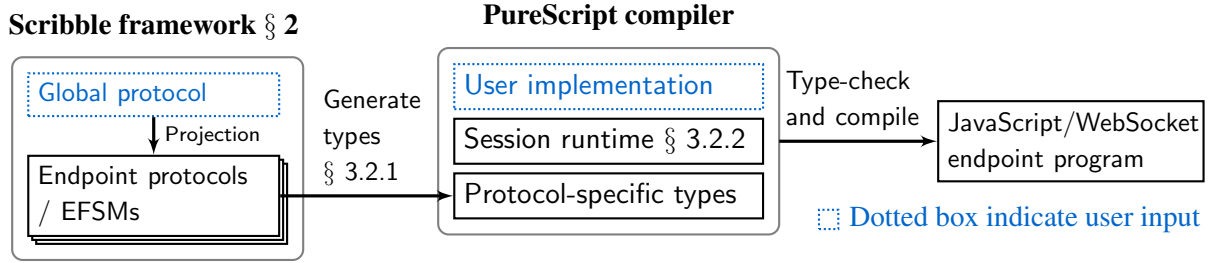
\begin{figure}
  \centering
  \input{fig-overview}
  \vskip -1.5em
  \caption{Overview of development work flow.}\label{fig:overview}
  \vskip -0.5em
\end{figure}

\figurename~\ref{fig:overview} presents our proposed type-safe web application
development work flow.
We implement our web application development framework in PureScript
on the top of the Scribble framework. 
%PureScript is a strongly-typed functional programming language designed
%to compile to readable JavaScript.
%It is heavily influenced by Haskell in its design and 
%syntax. In particular, PureScript
%supports type classes in the style of Haskell, which we employ in our work
%flow for type generation. %(\S~\ref{sec:gen}). 
%The language comes
%with a number of features that are unique due to its use cases in web
%development: 
%\begin{itemize}
%  \item Extensive libraries for web-based applications
%  \item Can integrate with existing JavaScript libraries easily
%  \item ...\JKnote{Add web-specific features of PureScript}
%\end{itemize}

%% file: fig-battleship.tex
\definecolor{browser-red}{HTML}{FF5148}
\definecolor{browser-yellow}{HTML}{FFBB00}
\definecolor{browser-green}{HTML}{18C934}
\colorlet{browser-title}{gray!40}
\begin{tikzpicture}[%
  browser/.style={draw=browser-title,thick,minimum width=5em,minimum height=5.3em},%
  msgpass/.style={>=latex',->},
  msglabel/.style={inner sep=1},
  initphase/.style={fill=white},
  p1phase/.style={fill=blue!10},
  p2phase/.style={fill=red!10},
]
  \node (svr) [draw,minimum width=4em,minimum height=5.3em] {\sffamily Server};
  \node (p1) [browser,left=3 of svr] {\sffamily Player 1};
  \fill [browser-title] (p1.north west) rectangle ($(p1.north east)+(0,0.2)$);
  \fill [browser-red] ($(p1.north west)+(0.1,0.1)$) circle (0.05);
  \fill [browser-yellow] ($(p1.north west)+(0.25,0.1)$) circle (0.05);
  \fill [browser-green] ($(p1.north west)+(0.4,0.1)$) circle (0.05);
  \node (p2) [browser,right=3 of svr] {\sffamily Player 2};
  \fill [browser-title] (p2.north west) rectangle ($(p2.north east)+(0,0.2)$);
  \fill [browser-red] ($(p2.north west)+(0.1,0.1)$) circle (0.05);
  \fill [browser-yellow] ($(p2.north west)+(0.25,0.1)$) circle (0.05);
  \fill [browser-green] ($(p2.north west)+(0.4,0.1)$) circle (0.05);
  \draw
    ($(p1.north)+(0.3,0.3)$)
      edge[<->] node[fill=white,inner sep=0.5] {\scriptsize\itshape WebSocket connection}
    ($(svr.north)+(-0.3,0.3)$);
  \draw
    ($(svr.north)+(0.3,0.3)$)
      edge[<->] node[fill=white,inner sep=0.5] {\scriptsize\itshape WebSocket connection}
    ($(p2.north)+(-0.3,0.3)$);
  \draw
    ($(p1.north east)+(0.1,-0.1)$)
      edge[msgpass,->] node [initphase,msglabel] {\scriptsize\sffamily Init(Config)}
    ($(svr.north west)+(-0.1,-0.1)$);
  \draw
    ($(p2.north west)+(-0.1,-0.1)$)
      edge[msgpass,->] node [initphase,msglabel] {\scriptsize\sffamily Init(Config)}
    ($(svr.north east)+(0.1,-0.1)$);
  \fill[blue!10] ($(p1.north east)+(0,-0.35)$) rectangle ($(svr.north west)+(0,-1.65)$);
  \fill[blue!10] ($(svr.north east)+(0,-0.35)$) rectangle ($(p2.north west)+(0,-1.65)$);
  \draw
    ($(p1.north east)+(0.1,-0.6)$)
      edge[msgpass,->] node [p1phase,msglabel] {\scriptsize\sffamily Attack(x,y)}
    ($(svr.north west)+(-0.1,-0.6)$);
  \draw
    ($(svr.north east)+(0.1,-1)$)
      edge[dashed,msgpass,->] node [p1phase,msglabel] {\scriptsize\sffamily Hit(x,y)}
    ($(p2.north west)+(-0.1,-1)$);
  \draw
    ($(svr.north west)+(-0.1,-1)$)
      edge[dashed,msgpass,->] node [p1phase,msglabel] {\scriptsize\sffamily Hit}
    ($(p1.north east)+(0.1,-1)$);
  \node at ($(p1.north east)+(0.5,-1.2)$) {\tiny\itshape or};
  \draw
    ($(svr.north east)+(0.1,-1.4)$)
      edge[dotted,msgpass,->] node [p1phase,msglabel] {\scriptsize\sffamily Miss(x,y)}
    ($(p2.north west)+(-0.1,-1.4)$);
  \draw
    ($(svr.north west)+(-0.1,-1.4)$)
      edge[dotted,msgpass,->] node [p1phase,msglabel] {\scriptsize\sffamily Miss}
    ($(p1.north east)+(0.1,-1.4)$);
  \node at ($(p2.north west)+(-0.5,-1.2)$) {\tiny\itshape or};
  \fill[red!10] ($(p1.north east)+(0,-1.65)$) rectangle (svr.south west);
  \fill[red!10] ($(svr.north east)+(0,-1.65)$) rectangle (p2.south west);
  \draw
    ($(p2.north west)+(-0.1,-1.9)$)
      edge[msgpass] node [p2phase,msglabel] {\scriptsize\sffamily Attack(x,y)}
    ($(svr.north east)+(0.1,-1.9)$);
  %\node at ($(svr.south east)+(1.25,0.1)$) {\scriptsize\sffamily\dots};
\end{tikzpicture}

%% file: fig-overview.tex
\definecolor{user}{rgb}{0,0.5,1}
\begin{tikzpicture}[%
    box/.style={draw,thick},
    next/.style={thick,>=latex',line width=1.2pt},%
    ep/.style={text width=8em,fill=white,box},%
    input/.style={densely dotted,user!80!black},
  ]
  \node [text width=8em,box,input]                  (g) {\small\sffamily Global protocol};
  \node [below=0.5 of g,ep,xshift=3pt,yshift=-3pt]     (e3) {\small\sffamily Endpoint protocols / EFSMs};
  \node [below=0.5 of g,ep,xshift=1.5pt,yshift=-1.5pt] (e2) {\small\sffamily Endpoint protocols / EFSMs};
  \node [below=0.5 of g,ep]                             (e) {\small\sffamily Endpoint protocols / EFSMs};
  \draw (g) edge[next,->] node [right] {\scriptsize\sffamily Projection} (e);
  \draw [rounded corners,thick,gray] ($(g.north west)+(-0.2,0.2)$) rectangle ($(e3.south east)+(0.2,-0.2)$);
  \node [right=2.3 of e,box,text width=10em] (api) {\sffamily Protocol-specific types};
  \node [above=0.1 of api.north east,anchor=south east,box,text width=10em] (rt) {\sffamily Session runtime \S~\ref{sec:runtime}};
  \node [above=0.1 of rt.north east,anchor=south east,box,input,text width=10em] (user) {\sffamily User implementation};
  \draw [rounded corners,thick,gray] ($(user.north west)+(-0.2,0.2)$) rectangle ($(api.south east)+(0.2,-0.2)$);
  \draw (e) edge[next,->] node [text width=3em,above] {\small\sffamily Generate types \S~\ref{sec:gen}} (api);
  \node [right=2.3 of rt,box,text width=8.5em] (out) {\small\sffamily JavaScript/WebSocket endpoint program};
  \draw ($(rt.east)+(0.2,0)$) edge [next,->] node [above,text width=5em] {\small\sffamily Type-check and compile} (out);
  \node [above=0.3 of user] {\bfseries PureScript compiler};
  \node [above=0.3 of g] {\bfseries Scribble framework \S~\ref{sec:scribble}};
  \node [right= of api.south east,input] (legend) {Dotted box indicate user input};
  \node [left=0 of legend.west,input,box] {};
\end{tikzpicture}

%% file: scribble.tex
\section{The Scribble protocol language}\label{sec:scribble}

Our development work flow extends Scribble~\cite{web:scribble,YHNN13},
a protocol specification language and code generation framework based on MPST.
Development starts by specifying the overall communication structure of the
target application as a \textit{global} protocol in Scribble, which is then
validated by the Scribble tool chain to ensure the protocol is well-formed.
Scribble
protocols are organised into \textit{modules} where each module contains
declaration of \textit{message payload types}, and one or more global protocol
definitions. 
%The bodies of protocol definitions are sequences of message passing
%and control flow statements.
We explain the syntax and structure of Scribble protocols using % a snippet of
the Battleship game Scribble protocol in
\lstlistingname~\ref{lst:battleship-game-protocol}.
First, we declare the module of the protocol
with the keyword \lstscribble{module} and the module name
(e.g.~line~\ref{line:module-game}).
It is followed by message type declaration statements using the
\lstscribble{type ... as} keywords. Messages in Scribble is written as
\lstscribble{Label(payloads)}, where \lstscribble{Label} is an identifying label
for the message, and \lstscribble{payloads} is a list of types for the
payloads of the message. For example, line~\ref{line:type} declares a new
payload type named \lstscribble{Location}, and specifies that the concrete type
corresponds to a PureScript data type \lstscribble{Game.Battleships.Location}.
It is later used on line~\ref{line:game-attack} in the protocol body as the
payload type of the \lstscribble{Attack(Location)} message.
By associating the message types used in a protocol with a concrete type from
the implementation language, messages specified in the protocol can be verified
against the implementation of the protocol, which we will discuss in more
details in the next section (\S~\ref{sec:gen}).

\begin{lstlisting}[%
  language=Scribble,style=ScribbleWithLN,basicstyle=\scriptsize\ttfamily,%
  firstline=1,lastline=22,%
  label=lst:battleship-game-protocol,%
  caption=Main body of the Battleships protocol.,%
]
module Game; /*<\label{line:module-game}>*/
type <purescript> "Config" from "Game.BattleShips" as Config; // Ship configuration /*<\label{line:battleship-config}>*/
type <purescript> "Location" from "Game.BattleShips" as Location; // Ship position /*<\label{line:type}>*/

global protocol Game(role Atk, role Svr, role Def) { /*<\label{line:game-protocol}>*/
    Attack(Location) from Atk to Svr; /*<\label{line:game-attack}>*/
    choice at Svr { // Svr knows if it's a hit /*<\label{line:game-choice:begin}>*/
        Hit(Location) from Svr to Atk; Hit(Location) from Svr to Def; /*<\label{line:game-hit:begin}>*/
        do Game(Def, Svr, Atk); /*<\label{line:game-hit:end}>*/
    } or {
        Miss(Location) from Svr to Atk; Miss(Location) from Svr to Def; /*<\label{line:game-miss:begin}>*/
        do Game(Def, Svr, Atk); /*<\label{line:game-miss:end}>*/
    } or {
        choice at Svr { /*<\label{line:game-end:begin}>*/
            Sunk(Location) from Svr to Atk; Sunk(Location) from Svr to Def;
            do Game(Def, Svr, Atk);
        } or {
            Winner() from Svr to Atk; Loser() from Svr to Def;
        } /*<\label{line:game-end:end}>*/
    } /*<\label{line:game-choice:end}>*/
}
\end{lstlisting}

Line~\ref{line:game-protocol} defines the global protocol \lstscribble{Game}
with three roles: an attacker (\lstscribble{Atk}), the server
(\lstscribble{Svr}), and a defender (\lstscribble{Def}).
In the body of the protocol, \lstscribble{Atk} first sends a coordinate to
attack (i.e.~a \lstscribble{Attack(Location)} message) to \lstscribble{Svr},
using a message passing statement on line~\ref{line:game-attack}.
After receiving the message, assuming \lstscribble{Svr} holds the coordinates
of the ship configurations of both the players, \lstscribble{Svr} will decide
whether the coordinate is a hit or a miss. Depending on the outcome, the
protocol will exhibit alternative behaviours. For instance, if it is a
\textit{hit}, \lstscribble{Svr} will send a \lstscribble{Hit} message to
\lstscribble{Atk} to notify it of the outcome, and also a \lstscribble{Hit}
message with the coordinate being attacked to \lstscribble{Def}; similarly if it
is a \textit{miss}, \lstscribble{Miss} message will be sent instead.
This is written in Scribble as a \lstscribble{choice} statement
(line~\ref{line:game-choice:begin}--\ref{line:game-choice:end}).
The syntax \lstscribble{choice at Svr} means the choice is being made at the
role \lstscribble{Svr}, and also that \lstscribble{Svr} will the first sender
of message in each of the possible branches: the \textit{hit} branch on
lines~\ref{line:game-hit:begin}--\ref{line:game-hit:end} or
the \textit{miss} branch on
lines~\ref{line:game-miss:begin}--\ref{line:game-miss:end}.
At the end of the \textit{hit} branch (line~\ref{line:game-hit:end}),
we use a \lstscribble{do} statement to recursively call the \lstscribble{Game}
protocol to move to the next round of the game. A \lstscribble{do} statement
includes role parameters to the protocol, where in this branch the current
\lstscribble{Atk} role and \lstscribble{Def} role continues as \lstscribble{Atk}
and \lstscribble{Def} respectively, such that the attacker can launch
consecutive attacks if they had been hit. However, in the last line of the
\textit{miss} branch (line~\ref{line:game-miss:end}), \lstscribble{Atk} and
\lstscribble{Def} are swapped, meaning that if there is a miss, then the current
defender gets a chance to attack in the next round of game.
Notice that in our protocol, we also describe a third branch in addition to the
\textit{hit} branch and \textit{miss} branch on
lines~\ref{line:game-end:begin}--\ref{line:game-end:end}. This branch describes
the situation when a battleship is sunk (i.e.~all coordinates of a ship are
hit). There are two possible outcomes, so we use a nested \lstscribble{choice}
statement to describe the two branches: the first branch is the game continues as a
\textit{hit} branch, but with a \lstscribble{Sunk} message in place of a
\lstscribble{Hit} message; the other branch is the endgame -- i.e.~all
battleships of one player are sunk -- in this branch, a winner is declared with
a \lstscribble{Winner} and \lstscribble{Loser} message,
and there are no \lstscribble{do} statements in this branch as the game ends
immediately with no need for a next round of game.

We use the Scribble tool chain to check that the Scribble protocol is well-formed.
For example, a \lstscribble{choice} statement is
well-formed only if the first message of all branches are sent by the choice
maker, otherwise the roles may be left in an inconsistent state.
A well-formed global protocol can then be \textit{projected} automatically into
an \textit{endpoint protocol} for each role in the protocol. An endpoint
protocol is a localised version of the global protocol which includes only the
interactions if they directly involve the target role. Scribble can represent an
endpoint protocol as an equivalent \textit{Endpoint Finite State Machine}
(EFSM), where the communication interactions of the protocol are represented by
transitions in the EFSM between the protocol states. We use the definitions of
EFSMs from Scribble in~\cite{HY2016} as a basis.

%% file: encoding.tex
\section{Endpoint programming with EFSMs}\label{sec:endpoint}

The EFSMs derived from the global protocol represent the local communication
behaviour at each of the endpoints, and is used as a guidance for developers to
implement their application endpoint.
To integrate the EFSMs in the user's programming work flow, our approach
interprets states and transitions in the EFSMs as (uninhabited) \textit{types} and type-class \textit{instances} in the target programming language.
We first present an encoding of the types of ESFM transition such as
\textbf{send} and \textbf{receive} as type classes (\S~\ref{sec:encoding}),
then we apply the encoding to generate types and instances for the user's application
(\S~\ref{sec:gen}).

\begin{figure}[t!]
  \tikzset{smallstate/.style={circle,draw,inner sep=0,text width=1.5em,text centered}}
  \begin{minipage}[t]{0.5\textwidth}
    \begin{lstlisting}[language=Haskell,style=hask,basicstyle=\ttfamily\footnotesize]
class Send r s$^!$ t a | s$^!$ $\leadsto$ t r a
    \end{lstlisting}
    \begin{tikzpicture}
      \node (ss) [smallstate,initial,initial text={}] {$s^!$};
      \node (st) [right=1.5 of ss,smallstate] {$t$}
       edge [<-,bend right]
       node [midway,below] {$!a$} (ss);
    \end{tikzpicture}
  \end{minipage}
  \begin{minipage}[t]{0.5\textwidth}
    \begin{lstlisting}[language=Haskell,style=hask,basicstyle=\ttfamily\footnotesize]
class Receive r s$^?$ t a | s$^?$ $\leadsto$ t r a
    \end{lstlisting}
    \begin{tikzpicture}
      \node (rs) [smallstate,initial,initial text={}] {$s^?$};
      \node (rt) [right=1.5 of rs,smallstate] {$t$}
       edge [<-,bend right]
       node [midway,below] {$?a$} (rs);
    \end{tikzpicture}
  \end{minipage}
  \begin{minipage}[t]{0.5\textwidth}
    \begin{lstlisting}[language=Haskell,style=hask,basicstyle=\ttfamily\footnotesize]
class Select r s$^!$ (ts :: RowList)
           | s$^!$ $\leadsto$ ts r
    \end{lstlisting}
    \begin{tikzpicture}
      \node (sels) [smallstate,initial,initial text={}] {$s^!$};
      \node (selt--) [right=3.2 of sels,smallstate] {$t^!_N$}
       edge [<-,bend left]
       node [midway,inner sep=1,fill=white] {$l_N$} (sels);
      \node (selt-) [right=2.2 of sels,smallstate] {$t^!_i$}
       edge [<-,bend right]
       node [midway,inner sep=1,fill=white] {$l_{i\in N}$} (sels);
      \node (selt) [right=1.2 of sels,smallstate,fill=white] {$t^!_1$}
       edge [<-]
       node [midway,inner sep=1,fill=white] {$l_1$} (sels);
      \draw (selt) edge [draw=none] node [midway] {\tiny\dots} (selt-);
      \draw (selt-) edge [draw=none] node [midway] {\tiny\dots} (selt--);
    \end{tikzpicture}
  \end{minipage}
  \begin{minipage}[t]{0.5\textwidth}
    \begin{lstlisting}[language=Haskell,style=hask,basicstyle=\ttfamily\footnotesize]
class Branch r r' s$^?$ (ts :: RowList)
           | s$^?$ $\leadsto$ ts r r'
    \end{lstlisting}
    \begin{tikzpicture}
      \node (bras) [smallstate,initial,initial text={}] {$s^?$};
      \node (brat--) [right=3.2 of bras,smallstate] {$t^?_N$}
       edge [<-,bend left]
       node [midway,inner sep=1,fill=white] {$l_N$}(bras);
      \node (brat-) [right=2.2 of bras,smallstate] {$t^?_i$}
       edge [<-,bend right]
       node [midway,inner sep=1,fill=white] {$l_{i\in N}$} (bras);
      \node (brat) [right=1.2 of bras,smallstate,fill=white] {$t^?_1$}
       edge [<-]
       node [midway,inner sep=1,fill=white] {$l_1$} (bras);
      \draw (brat) edge [draw=none] node [midway] {\tiny\dots} (brat-);
      \draw (brat-) edge [draw=none] node [midway] {\tiny\dots} (brat--);
    \end{tikzpicture}
  \end{minipage}
  \caption{Send (top left), Receive (top right), Label selection (bottom left),
  and Label branching (bottom right) transition type classes and their state
  diagrams.}\label{fig:typeclasses}
  \vskip -1em
\end{figure}
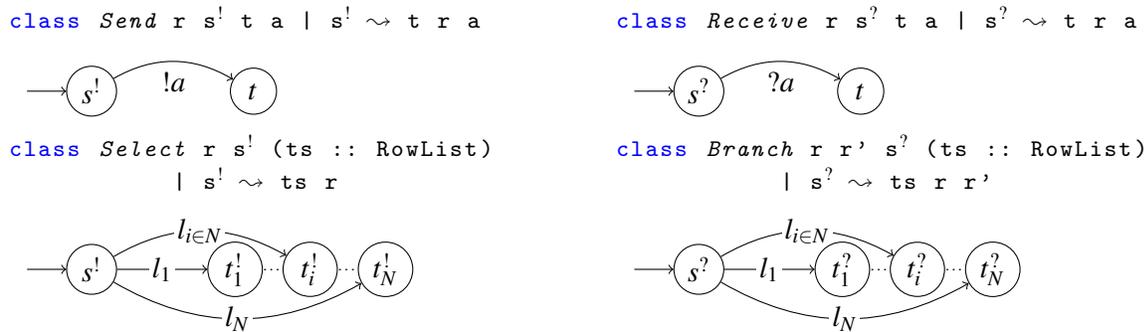

\subsection{Transitions as Type Classes}\label{sec:encoding}

A key contribution of this paper is our encoding of EFSMs
%generated
from a
Scribble protocol into type checking constraints with
\textit{multi-parameter type classes} (MPTCs).
Type classes~\cite{typeclass} were introduced to allow functions to be overloaded, where a type is polymorphic but constrained to be an \textit{instance} of the class. Examples of common type classes include \texttt{Eq a}, \texttt{Show a} and \texttt{Monoid a}.
MPTCs~\cite{jones:mptc} extend this by
allowing more than one type parameter in a type class definition. 
Combined with \textit{functional dependencies} between type parameters, which describe that a subset of the parameters uniquely determine another, it is possible to encode \textit{relations} between types. 

%Unspecified types are solved during type class instance resolution.

By representing kinds of transitions as MPTCs, it is possible to both constrain the usage of the corresponding functions and bring into context additional information about the transition, such as the next state, the type of the value communicated and the role involved.
Our encoding exploits the properties of EFSMs derived from a well-formed
protocol by the Scribble toolchain. The properties guaranteed by EFSMs as noted
in~\cite{HY2016} include: (1) there is exactly one initial state; (2) there is
at most one terminal state; and (3) every state in an EFSM is one of
three kinds: an \textit{output (resp. input) state}  where every
transition is an output (resp. input) or a terminal state.

We first consider EFSM transitions where the current state has only a single
transition (hence a single successor state).
\figurename~\ref{fig:typeclasses} (top row) illustrates the type class definitions for
output and input states.
The type parameters $s$, $t$, and $a$ are the type representing the
\textit{current} and the \textit{successor} state of the transition, and the
message \textit{payload type} of the output and input action. Our current state \textit{s} determines all other parameters, which the functional dependency $s$ $\leadsto$ $t$ $r$ $a$ describes.
A similar pair of transition type classes exist for \lsthask{Connect} and
\lsthask{Disconnect} for establishing and terminating connections
(i.e.~connection actions from~\cite{HY2017}) respectively
but omitted here due to space constraints.

For output $S^!$ and input states $S^?$ in the EFSM which have multiple
transitions and successor states (i.e.~branching and selection), each of the
transitions can be identified by their message payload \textit{label},
used for determining the selected branch between the sender and the
receiver. Our encoding makes the label selection explicit, by splitting each of
the output (or input) transitions into two parts: a label send (resp.~receive),
then the actual output (resp.~input) action. A new intermediate state $T$ is
introduced between the two transitions and acts as the output (or input) state
for the actual output (resp.~input) transition for each branch, such that
transitions of $T$ can be encoded into \texttt{Send} or \texttt{Recv} type
classes above.
The original multi-transition states are no longer output and input states after
the transformation, as the transitions now perform label send and receive
instead of output and input actions. We encode the set of \textit{label send}
and \textit{receive} transitions from the same state as \texttt{Branch}
and \texttt{Select} type classes, as depicted in
\figurename~\ref{fig:typeclasses} (bottom row) for a type-safe mapping between
the labels and the chosen branches.
The type parameter $s$ is the initial state of the transition, and
$ts$ is a \textit{row list} of tuples $(l_i, t_i)_{i\in{}|ts|}$
containing type-level string label $l_i$, and its
corresponding continuation state $t_i$ for the branch. Instances of $ts$
are used to express the finite number of possible branches in the EFSM,
such that branches with undefined labels in the EFSM cannot be used.
Similar to the \texttt{Send} and \texttt{Recv} type classes, the type classes
are annotated with their functional dependencies, indicating that instances of
the initial state $s$ uniquely determines $ts$.

%% file: impl.tex
\subsection{Implementation}
\label{sec:implementation}
\subsubsection{Types generation}\label{sec:gen}

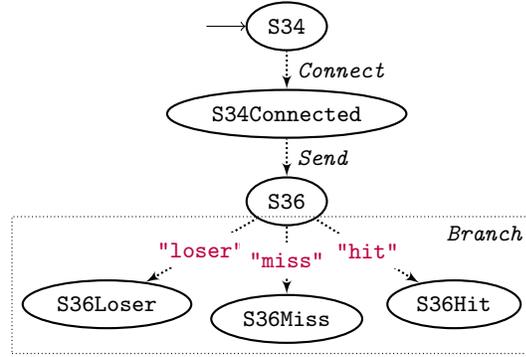
\begin{figure}[t]
\begin{lstlisting}[%
  language=Haskell,style=hask,basicstyle=\ttfamily\scriptsize,
]
import Game.BattleShips (Config, Location) --/*\tikz[remember picture]{\coordinate (efsm-here);}*/--
...
data Init = Init Config
...
foreign import data BattleShips :: Protocol
foreign import data P2 :: Role
instance roleNameP2 :: RoleName P2 "P2"
...
foreign import data S34 :: Type
foreign import data S34Connected :: Type
foreign import data S36 :: Type
...
instance initialP2 :: Initial P2 S34
instance terminalP2 :: Terminal P2 S35
instance connectS34 :: Connect P2 GameServer S34 S34Connected
instance sendS34 :: Send GameServer S34Connected S36 Init
instance branchS36 :: Branch P2 GameServer S36
  (Cons "loser" S36Loser (Cons "miss" S36Miss (Cons "hit" S36Hit Nil)))
...
\end{lstlisting}
\begin{tikzpicture}[
    st/.style={ellipse,draw,thick},
    tr/.style={>=latex',densely dotted,thick},
    brlab/.style={fill=white},
    remember picture,overlay,
  ]
  \node [initial,st,initial text=,right=9em of efsm-here,anchor=west] (s34)
  {\footnotesize\tt S34};
  \node [st,below=0.5 of s34] (s34c) {\footnotesize\tt S34Connected}
   edge [<-,tr]
   node [right] {\footnotesize\ttfamily\itshape Connect} (s34);
  \node [st,below=0.5 of s34c] (s36) {\footnotesize\tt S36}
   edge [<-,tr]
   node[right] {\footnotesize\ttfamily\itshape Send} (s34c);
  \node [st,below left=0.9 and 1.2 of s36] (s36l) {\footnotesize\tt S36Loser}
   edge [<-,tr]
   node [brlab] {\footnotesize\lsthask{"loser"}} (s36);
  \node [st,below=0.9 of s36] (s36m) {\footnotesize\tt S36Miss}
   edge [<-,tr]
   node [brlab] {\footnotesize\lsthask{"miss"}} (s36);
  \node [st,below right=0.9 and 1.2 of s36] (s36h) {\footnotesize\tt S36Hit}
   edge [<-,tr]
   node [brlab] {\footnotesize\lsthask{"hit"}} (s36);
  \node [fit=(s36l)(s36m)(s36h)(s36.south),draw,densely dotted] (s36fit) {};
  \node [below=0 of s36fit.north east,anchor=north east] {\footnotesize\ttfamily\itshape Branch};
\end{tikzpicture}
  \vskip -2em
  \caption{Example fragment of generated types and its corresponding EFSM.}\label{lst:generated}
  \vskip -1em
\end{figure}

Given a valid Scribble protocol, our framework generates from the corresponding
EFSM a new data type to represent each state, and the transitions
%in the EFSM
between the states are instantiated as type instances of type classes described
in \S~\ref{sec:encoding}. The types generated by the framework is a static
guidance for users to use the \textbf{session runtime}
%provided
to perform communication in way that conforms to the input protocol.
A fragment of the EFSM and its corresponding generated types for the Battleships
protocol is given in \figurename~\ref{lst:generated}.
Message payload types are imported to the PureScript module from the specified
path in the protocol, and are defined by the user.
%Data types for all message payloads in the protocol are generated.
For example, the \lsthask{Config} type in \figurename~\ref{lst:generated}
corresponds to the \lstscribble{type} declaration of the same name in the
protocol (line~\ref{line:battleship-config} in
\lstlistingname~\ref{lst:battleship-game-protocol}), imported from user-defined
\texttt{Game.Battleships.Config}.
\texttt{Role} and \texttt{RoleName} are also generated from the protocol, where
the latter provides metadata (a symbol representation) of a role, and is used
by the runtime as a key for accessing communication channels.
Finally, the type class instances \lsthask{initialP2} and \lsthask{terminalP2}
are the initial and terminal transitions (no predecessor and successor states
respectively);
\lsthask{connectS34}, \lsthask{sendS34} and \lsthask{branchS36} are normal
transitions for \texttt{connect}, \texttt{send} and \texttt{branch} actions.
For each \texttt{branch} transition, a \texttt{RowList} (a way to inductively
represent a row type) is used to describe labels that can be chosen from and
their corresponding successor, e.g.~\lsthask{loser} label corresponds to
the \lsthask{S36Loser} successor state.

\subsubsection{Session runtime}\label{sec:runtime}\label{sec:combinators}
The session runtime is a library of communication combinators that can only be used to construct correct protocol implementations.
These are:
\lsthask{connect}/\lsthask{disconnect} (for managing connections between participants),
\lsthask{send}/\lsthask{receive} (for point-to-point message passing),
\lsthask{choice}/\lsthask{branch} (for branching and selection) and \lsthask{session} (for running the session).

\begin{lstlisting}[%
  language=Haskell,style=hask,basicstyle=\scriptsize\ttfamily,%
]
newtype Session m c i t a = Session ((Channels c i) -> m (Tuple (Channels c t) a))

bind :: Monad m => Session m c i s a -> (a -> Session m c s t b) -> Session m c i t b
\end{lstlisting}

A \lsthask{Session} (above) is a continuation that consumes a \lsthask{Channel} type, 
indexed by an \textit{initial} state channel type \lsthask{s},
and (effectfully) produces a channel in a \textit{terminal} state \lsthask{t}
with a result \texttt{a}. This definition is not exported outside of the module, so the only way to construct a \lsthask{Session} is through using one of the communication combinators. We can compose two sessions where the terminal state of the first is the initial of the second, resulting in a session that starts at initial state of the first and ends in the terminal state of the second using (an indexed~\cite{ixmonad}) \lsthask{bind}. Given a session whose initial and terminal states match that of a protocol (provided by the \texttt{Initial} and \texttt{Terminal} constraints) using \lsthask{session} we can `run' it (to produce a monadic value). The user is not required to provide concrete states, as the type checker can determine them by solving the constraints.

%We observe that \lsthask{Session} is a specialisation of \lsthask{IxStateT} -- the type changing monad state transformer -- and could be written using it.
In \figurename~\ref{fig:runtime} we provide the types of some session combinators. \lsthask{send} can be read as ``given a value \texttt{a} can be encoded as JSON and there is a role \texttt{r} you can send the value to, by transitioning from state \texttt{s} to \texttt{t}, then it will produce the session that starts at \texttt{s} and terminates at \texttt{t} producing the \texttt{Unit} value''. \lsthask{receive} is similar except instead the session \textit{produces} the value received. \lsthask{choice} can be read as ``given a record providing continuations for each of the branches \texttt{ts} from the state \texttt{s} that reach the terminal state \texttt{u} producing a value \texttt{a}, then it will produce a session that starts at \texttt{s} and terminates at \texttt{u} producing \texttt{a}''. The runtime will then select the appropriate continuation based on the message received.
%\JKnote{I'm not sure where to put this}
%
%The \texttt{Aff} monad describes asynchronous effects and can be thought of as
%similar to \texttt{IO} in Haskell (without preemption, as PureScript is
%evaluated by the single threaded JavaScript event loop).
The \texttt{MonadAff}
class used by all of the combinators can be thought of like \texttt{MonadIO} in Haskell and allows
%\texttt{Aff}
the asynchronous communication effects
to be \textit{lifted} into an arbitrary monad stack.
A concrete example of their use in our framework is shown in \S~\ref{sec:lifting}.

\myparagraph{Implementing branching}
In our \lsthask{choice} combinator we need to decode the JSON, which we will apply the continuation to, however we only know what type it should be decoded to at runtime based on the message we receive.
One solution would be to manually decode the JSON in the continuation, however this is unsatisfactory as the runtime library should consistently abstract this. We solve this by inserting \texttt{Receive} states, which a branch continuation must begin with -- this works because we statically know all possible types the value could be and enumerate them all. This is like receiving a label selecting the branch, followed by the value, however in practice a single message is communicated. The runtime selection relies on the JSON encoding of message data type, however as these are generated this is safe.%[1mm]

\begin{figure}[t!]
\begin{minipage}[t]{0.44\linewidth}
\begin{lstlisting}[%
  language=Haskell,style=hask,basicstyle=\scriptsize\ttfamily,%
  label=lst:runtime-session,%
]
session :: forall r c p s t m a.
     Transport c p
  => Initial r s
  => Terminal r t
  => MonadAff m
  => Proxy c
  -> Role r
  -> Session m c s t a
  -> m a
\end{lstlisting}
\end{minipage}
  \begin{minipage}[t]{0.56\linewidth}
\begin{lstlisting}[%
  language=Haskell,style=hask,basicstyle=\scriptsize\ttfamily,%
  label=lst:runtime-choice,%
]
choice :: forall r r' rn c s ts u funcs row m p a.
     Branch r r' s ts
  => RoleName r' rn
  => IsSymbol rn
  => Terminal r u
  => Transport c p
  => Continuations (Session m c) ts u a funcs
  => ListToRow funcs row
  => MonadAff m
  => Record row -> Session m c s u a
\end{lstlisting}
\end{minipage}

\noindent
  \begin{minipage}[t]{0.5\linewidth}
\begin{lstlisting}[%
  language=Haskell,style=hask,basicstyle=\scriptsize\ttfamily,%
  label=lst:runtime-send,%
]
send :: forall r rn c a s t m p.
     Send r s t a
  => RoleName r rn
  => IsSymbol rn
  => Transport c p
  => EncodeJson a
  => MonadAff m
  => a -> Session m c s t Unit
\end{lstlisting}
\end{minipage}
  \begin{minipage}[t]{0.5\linewidth}
\begin{lstlisting}[%
  language=Haskell,style=hask,basicstyle=\scriptsize\ttfamily,%
  label=lst:runtime-recv,%
]
receive :: forall r rn c a s t m p.
        Receive r s t a
     => RoleName r rn
     => IsSymbol rn
     => Transport c p
     => DecodeJson a
     => MonadAff m
     => Session m c s t a
\end{lstlisting}
\end{minipage}
\caption{Sample types and primitives in the session runtime.}\label{fig:runtime}
\end{figure}

\myparagraph{Transport abstraction}
In our example communication is performed over WebSocket connections, however this can be generalised to a reliable, order-preserving asynchronous channels. The type variable \texttt{c} in \texttt{Session} and \texttt{Channel} allows this parameterisation. Communication in our runtime is implemented through abstract primitives defined in a \texttt{Transport} type class, allowing library users to provide their own \textit{transport} layer. With this abstraction we have additionally implemented support for using \texttt{AVar}'s (similar to Haskell's \lsthask{MVar}) to perform communication locally through shared memory.\\[1mm]
\myparagraph{Linear usage of channel}
Through our careful choice of library design, inspired by existing work using Indexed Monads~\cite{ixmonad}, channels are not directly accessed by the programmer. This means \textit{reuse} is impossible and by requiring continuation to a terminal state to `run' a session, \textit{use} is guaranteed. A faulty runtime implementation could still violate this, however this only has be to verified once by the library author rather than the user.

%% file: casestudy.tex
\section{Case study}
We present an implementation of the Battleship running example in PureScript
using our framework. The full implementation can be found
in~\cite{gh:scribble-battleships}.
Most web frameworks (e.g.~React or Halogen) are event driven, where
an \emph{update} function handles events fired by user interaction. There is no
knowledge however about the order in which events are received, or that only a
subset of events are possible in a given state (e.g.~if currently a 
button is disabled). This means that it is not possible to have a session
that spans more than one event in an event driven framework,
while still preserving static linearity guarantees by construction.
We therefore use the Concur UI framework that constructs UIs sequentially,
which is a perfect fit for inherently sequential sessions.\\[1mm]
\myparagraph{Widgets}
%https://github.com/ajnsit/concur-documentation/blob/master/README.md cite (maybe we also want to reword the following)
Concur is built around composing Widgets. A Widget is something that has a view, can internally update in response to some events, and will return some value.
Consider the following snippet of code, which defines
a \lsthask{play} button that displays \lsthask{"Play game"} and returns once is has been pressed.
\begin{lstlisting}[language=Haskell,style=hask,basicstyle=\ttfamily\footnotesize]
play :: forall a. Widget HTML Unit
play = button' [unit <$\mathtt{\$}$ onClick] [text "Play game"]
\end{lstlisting}

\noindent
We can then sequence this button with the text \lsthask{"Game over!"}, which will be displayed \textit{after} the button has been pressed. Note that the type \lsthask{"forall a. a"} means this widget can never return (as it is impossible to produce a value of this type) and so will be displayed forever.
\begin{lstlisting}[language=Haskell,style=hask,basicstyle=\ttfamily\footnotesize]
example :: forall a. Widget HTML a
example = do
  play
  text "Game over!"
\end{lstlisting}
\myparagraph{Lifting widgets}\label{sec:lifting}
Our runtime combinators are parameterised over any \lsthask{MonadAff}%
\footnote{The \texttt{MonadAff} class is an asynchronous effect monad, in some respects similar to \texttt{MonadIO} in Haskell.}
so that we can pick \lsthask{Widget HTML} from the Concur UI framework. \lsthask{lift} lets us \textit{lift} a widget into a session, which produces a value while remaining in the same session state.
\begin{lstlisting}[language=Haskell,style=hask,basicstyle=\ttfamily\footnotesize]
lift :: forall c i f a. Functor f => f a -> Session f c i i a
\end{lstlisting}

\noindent
We build a widget that plays the game as Player 1, with interleaved sessions and user input. We begin by first connecting to the \lsthask{GameServer}, followed by running the \lsthask{setupGameWidget} to allow the user to place their ship. This configuration is then sent to the \lsthask{GameServer} and the player attacks. (Note: we use PureScript's support for rebinding \lsthask{bind} and \lsthask{pure} in a \lsthask{do} block).
\begin{lstlisting}[language=Haskell,style=hask,basicstyle=\ttfamily\footnotesize,mathescape=false]
battleShipsWidgetP1 url :: URL -> Widget HTML Unit
battleShipsWidgetP1 = session 
  (Proxy :: Proxy WebSocket)
  (Role :: Role P1) $ do
    connect (Role :: Role GameServer) url
    config <- lift setupGameWidget
    send $ Init config
    let pb = mkBoard config
    let ob = mempty :: Board OpponentTile
    attack pb ob
\end{lstlisting}
    
%    As \texttt{Widget v} is a Monad, our Session with the widget embedded can be 
%
%    \begin{lstlisting}[language=Haskell,style=hask,basicstyle=\ttfamily\footnotesize]
%instance sessionIxMonad :: Monad m => IxMonad (Session m c)
%    \end{lstlisting}
%
%
%\begin{lstlisting}[language=Haskell,style=hask,basicstyle=\ttfamily\footnotesize]
%
%instance sessionMonadAff :: MonadAff m => MonadAff (Session m c i i) where
%  liftAff aff = Session \c -> map (Tuple c) (liftAff aff)
%    \end{lstlisting}

\myparagraph{Limitations}
It is difficult to extract common `source code' to a single function, as its type is dependent on where it is used in the protocol. This can result in duplicated identical code, but with a different type signature. Technically it is possible to write out the state polymorphic version, however you would need to chain all the FSM constraints required by the runtime combinators which is tedious. This is a fundamental limitation of the FSM representation compared to actually embedding session types in the language. 
% Removing because we are already providing an abstract interface with the type classes, but still have this issue.
%In~\cite{HY2016}, some of these concerns was addressed by abstract interfaces
%for I/O states, where Java interfaces representing I/O are passed as parameters
%instead of concrete types. It is our future work to consider equivalent
%techniques in our framework.
Additionally when splitting an implementation into multiple functions the user will need to provide types for the initial and terminal states of each sub-protocol. For the same practical reasons, this will be a concrete generated type (e.g. \lsthask{S20}), meaning any changes to the Scribble protocol will require the type to be updated manually.\\

% Maybe include that websocket connections can only be received by a server/cannot be accepted by the browser.

%% file: related.tex
\section{Related work}
\myparagraph{Scribble-based code generation}
Code generation from Scribble is an effective way of applying
multiparty session types in mainstream programming languages.

Notably, Scribble-Java~\cite{HY2016,HY2017} was the earliest work to propose
hybrid session verification, by generating Java API from Scribble to statically
type check user's I/O action usages against the generated APIs, combined with
runtime checking of linear channel usages. The work focussed on desktop
applications and only support TCP and HTTP as the communication transport.
Scribble-Scala~\cite{SDHY2017} uses Scribble to generate Scala APIs that use
the lchannel library through a linear decomposition of multiparty session types.
The implementation uses the Akka actor framework and supports all transport
abstractions provided by Akka.
Neykova et al.~\cite{NHYA2018} implemented in F\# a session type provider
to support on-demand compile-time protocol by generating protocol-specific
.NET types from Scribble.
Pabble~\cite{NCY2015} generates C/MPI skeleton code
from parameterised Scribble for correct-by-construction role-parameterised MPI
parallel programming but do not use types to check for conformance.
The approach was revisited in~\cite{CHJNY2019} using a new
distributed formalisation of parameterised Scribble and applied to Go in the
Scribble-Go toolchain, with support for TCP and in-process shared memory
transport.
StMungo~\cite{Kouzapas:2016:TPM,stmungo} uses Scribble to generate typestate
definitions for static type checking of communication protocols in Java.

This is the first work that applies the session-based API generation
approach to WebSocket transport, and targets JavaScript applications for
the web.\\[1mm]
\myparagraph{Session types in functional languages with linearity}
There are many approaches to embed session types natively in advanced
type systems found in functional programming languages. The most challenging
aspects of the embedding to support full session type verification is
ensuring linear usage of channel resources.
A more comprehensive survey of session types with linearity in functional
languages can be found in~\cite{OY2017}, we highlight a few works that are
closely related to our approach.
The Links web programming language~\cite{web:links,LM2017} 
is a functional language designed
for tierless web programming, and recently adds a support for \textit{binary}
session types in the style of GV~\cite{gv}, and has an extension to support
linear types. Adding support for multiparty session types would
require to extend the core calculus of the language.
Padovani's FuSe~\cite{Padovani17A,MP2017} infers and type checks usage of
binary session-based communication in OCaml, and uses a hybrid static/dynamic
linearity check similar to the hybrid verification approach in~\cite{HY2016}.
Session-ocaml~\cite{IYY2017} implements 
session types in OCaml with lenses overcoming a linearity issue but
only treats binary session types.

To the best of our knowledge, this work is the first which implements 
(i.e. generates code from) \emph{multiparty session types} with 
\emph{fully static} linearity guarantees.

\section{Conclusion and Future work}
%\myparagraph{Future work} We see no reason why this work couldn't be applied to Haskell, although without row types the ergonomics would be slightly worse. By writing a typechecker plugin~\cite{Gundry:2015:TPU:2804302.2804305} we could automate/remove the code generation process and also provide better error messages by directly traversing the FSM during type checking.
%While WebSockets are bidirectional, a connection can only be opened by the browser, which prevents use in peer-to-peer browser communication. WebRTC could be explored as a solution for this use case.\\
%%\noindent
%In our work we treat user interaction as a source of input separate to the session, however what if you could describe interaction as a session, treating widgets as roles? Could we benefit from properties that well typed protocols provide (i.e. progress)? We believe this is an interesting design space where eventually the need for (a subset of) UI testing could be replaced by type checking.
%
We have presented a type-safe web application development work flow
following the MPST framework, by encoding Endpoint FSMs as type classes
and generating PureScript code from the Endpoint FSMs.

Future work include applying the approach to Haskell, although we suspect
the ergonomics would be slightly worse without native row types support.
The code generation process can be automated by
a typechecker plugin~\cite{Gundry:2015:TPU:2804302.2804305}, and also provide
better error message by directly traversing the FSM during type checking.
While WebSockets are bidirectional, a connection can only be opened by the
browser, which prevents use in peer-to-peer browser communication. WebRTC could
be explored as a solution for this use case.
In our work we treat user interaction as a source of input separate to the
session, however by describing interaction as a session, treating
widgets as roles, we may benefit from properties that well typed protocols
provide (i.e. progress). We believe this is an interesting design space where
eventually the need for (a subset of) UI testing could be replaced by type
checking.